# Dative epitaxy of commensurate monocrystalline covalent-van der Waals moiré supercrystal

*Mengying Bian[1,2]†, Liang Zhu[3]†, Xiao Wang[4], Junho Choi[5], Rajesh V. Chopdekar[6], Sichen Wei[7], Lishu Wu[8], Chang Huai[2], Austin Marga[2], Qishuo Yang[3], Yuguang C. Li[9], Fei Yao[7], Ting Yu[8], Scott A. Crooker[5], Xuemei M Cheng[4], Renat F. Sabirianov[10], Shengbai Zhang[11], Junhao Lin[3]\*, Yanglong Hou[1]\* & Hao Zeng[2]\**

[1]Beijing Key Laboratory for Magnetoelectric Materials and Devices, Beijing Innovation Center for Engineering Science and Advanced Technology, School of Materials Science and Engineering, Peking University, Beijing, China

[2]Department of Physics, University at Buffalo, State University of New York, Buffalo, NY, USA

[3]Department of Physics and Shenzhen Key Laboratory of Advanced Quantum Functional Materials and Devices, Southern University of Science and Technology, Shenzhen, China

[4]Physics Department, Bryn Mawr College, Bryn Mawr, PA, USA

[5]National High Magnetic Field Laboratory, Los Alamos National Laboratory, Los Alamos, NM, USA

[6]Advanced Light Source, Lawrence Berkeley National Laboratory, Berkeley, CA, USA

[7]Department of Materials Design and Innovation, University at Buffalo, The State University of New York, Buffalo, NY, USA

[8]Division of Physics & Applied Physics, School of Physical and Mathematical Sciences, Nanyang Technological University, Singapore

[9]Department of Chemistry, University at Buffalo, The State University of New York, Buffalo, NY, USA

[10]Department of Physics, University of Nebraska-Omaha, Omaha, NE, USA

[11]Department of Physics, Rensselaer Polytechnic Institute, Troy, NY, USA

†These authors contributed equally: Mengying Bian, Liang Zhu

*Corresponding author. e-mail: haozeng@buffalo.edu; hou@pku.edu.cn; linjh@sustech.edu.cn

**Abstract**

Realizing van der Waals (vdW) epitaxy in the 80's represents a breakthrough that circumvents the stringent lattice matching and processing compatibility requirements in conventional covalent heteroepitaxy. However, due to the weak vdW interactions, there is little control over film qualities by the substrate. Typically, discrete domains with a spread of misorientation angles are formed, limiting the applicability of vdW epitaxy. Here we report the epitaxial growth of monocrystalline, covalent $Cr_5Te_8$ 2D crystals on monolayer vdW $WSe_2$ by chemical vapor deposition, driven by interfacial dative bond formation. The lattice of $Cr_5Te_8$, with a lateral dimension of a few ten microns, is fully commensurate with that of $WSe_2$ via $3 \times 3$ ($Cr_5Te_8$)/$7 \times 7$ ($WSe_2$) supercell matching, forming a single-crystalline moiré superlattice. Our work has established a conceptually distinct paradigm of thin film epitaxy termed "dative epitaxy", which takes full advantage of covalent epitaxy with chemical bonding for fixing the atomic registry and crystal orientation, while circumventing its stringent lattice matching and processing compatibility requirements; conversely, it ensures the full flexibility of vdW epitaxy, while avoiding its poor orientation control. $Cr_5Te_8$ 2D crystals grown by dative epitaxy exhibit square magnetic hysteresis, suggesting minimized interfacial defects that can serve as pinning sites.

## 1. Introduction

Two dimensional (2D) heterostructures obtained by stacking van der Waals (vdW) layers have attracted intense interest for fundamental research and applications in electronics[1], optoelectronics[2], spintronics[3], and valleytronics[4]. In particular, moiré superlattices achieved by aligning or twisting individual 2D layers offer an additional degree of freedom for manipulating the electronic structure. It is well known that correlated insulating states, superconductivity, magnetism and topological quantum states can emerge in twisted bilayer graphene [5], graphene/hexagonal boron nitride [6] and transition metal dichalcogenide (TMD) moiré superlattices[7]. Moiré superlattice exciton states and interlayer valley excitons were also observed in TMD heterostructures such as $WSe_2/WS_2$ and $MoSe_2/WSe_2$, respectively [8]. However, the conventional exfoliation and stacking approach lacks scalability for practical applications.

Recently, 2D vdW heterostructures have been realized by chemical vapor deposition (CVD), such as graphene/hBN[9], and transition metal dichalcogenide (TMD) heterostructures (e.g., $WS_2/MoS_2$, $SnS_2/MoS_2$, and $NbTe_2/WSe_2$)[10]. VdW epitaxy overcomes the constraints of lattice matching and processing compatibility requirements in conventional covalent heteroepitaxy [11]. It is particularly suitable for synthesizing 2D heterostructures owing to the atomically smooth and dangling bond-free vdW surface. However, because vdW surfaces are chemically inactive, chemical or plasma treatment may be needed to facilitate nucleation, which leads to a defective interface[12]. Moreover, limited success has been achieved for vdW epitaxy of continuous thin films of materials with 3D crystal structures on vdW substrates. Instead, discrete nanowires or domains with a spread of misorientation alignment were typically obtained[13]. This is because the weak vdW interaction and the resulting energy landscape as a function of the in-plane orientation angle may not exhibit clearly defined minima, required for high-quality epitaxy[13-14].

In this work, we report dative epitaxy of a high-quality single-crystalline layer of 3D-bonded material on a 2D vdW template with large lattice mismatch. This unexplored regime of epitaxy exploits the bonding duality at the interface to realize epitaxial growth: the weak vdW interactions allow facile surface diffusion of precursor molecules for the growth of large area continuous layers, while the formation of dative bond (a special covalent bond where the bonding electrons derive from the same atom) fixes atomic registry and crystal orientation. Specifically, we focus on the 2D heterostructure of $Cr_5Te_8/WSe_2$, where $Cr_5Te_8$ is a non-vdW ferromagnet that can be considered as Cr atoms self-intercalated in-between the $CrTe_2$ layers[15] (as shown schematically in Figure S1a, supporting information). First, a large-scale $WSe_2$ monolayer millimeter in size was grown on a sapphire or $Si/SiO_2$ substrate by CVD. Highly aligned 2D $Cr_5Te_8$ crystals with thicknesses down to a single unit cell and yet sizes of tens of microns were then achieved by dative epitaxy, with $WSe_2$ as the vdW template. As a result, a globally commensurate, monocrystalline $3 \times 3/7 \times 7$ $Cr_5Te_8/WSe_2$ moiré supercrystal is achieved, which differs from conventional moiré superlattices with spatially varying rigid moiré patterns[16] or local commensurate domain reconstruction[17]. Three decades after the realization of vdW epitaxy[11a], our work redefines the scope and applicability of epitaxy with unprecedented opportunities for applications.

## 2. Results and Discussion

A schematic of the growth process of 2D $Cr_5Te_8/WSe_2$ moiré superlattices is shown in **Figure 1**a, which consists of monomer adsorption, desorption, and surface diffusion. Further details of the growth process are shown in Methods. CVD grown monolayer $WSe_2$ with lateral dimensions of $100 - 2,000$ μm, used as templates for the growth of $Cr_5Te_8/WSe_2$ heterostructures, are shown in Figure 1b, c[18]. A typical optical microscope image of the heterostructures is shown in Figure 1d. Strikingly, all the $Cr_5Te_8$ crystals grown on a single monolayer $WSe_2$ are self-aligned, with one of their edges oriented either parallel or at a 60° angle to one of the edges of WSe$_2$, in contrast to the randomly oriented Cr$_5$Te$_8$ crystals grown directly on sapphire (Figure S2a, supporting information). This strongly suggests that the Cr$_5$Te$_8$ crystals grow epitaxially on WSe$_2$. While monolayer WSe$_2$ are randomly oriented, the orientations of the Cr$_5$Te$_8$ crystals align with individual WSe$_2$ crystals, suggesting the dominant role of monolayer WSe$_2$ in the epitaxial growth. Therefore, such vdW templates also allow the synthesis of highly oriented 2D Cr$_5$Te$_8$ crystals independent of substrates, as evidenced by samples grown on amorphous SiO$_2$ substrates (Figure S2g, h, supporting information).

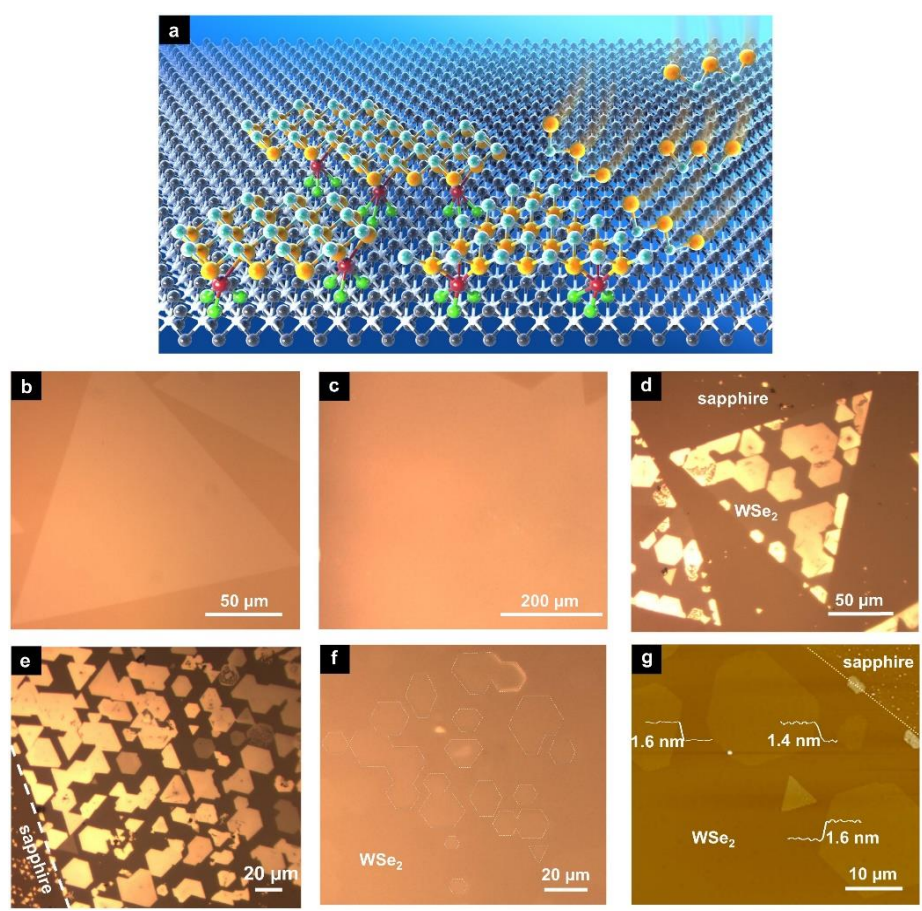

**Figure 1. Epitaxial growth process of Cr$_5$Te$_8$/WSe$_2$ heterostructures, optical and atomic force microscope images of WSe$_2$ and Cr$_5$Te$_8$/WSe$_2$ heterostructures. a,** A schematic of the epitaxial growth processes of 2D Cr$_5$Te$_8$/WSe$_2$ heterostructures, showing monomer adsorption, desorption, and diffusion. It also shows chemical bonding between interfacial Cr (red) and Se atoms (green). **b, c,** Optical microscope images of monolayer WSe$_2$ of sizes of ~ 200 μm (**b**) and ~ 1 mm (**c**); **d,** 2D Cr$_5$Te$_8$/WSe$_2$ heterostructures; **e, f** Highly aligned Cr$_5$Te$_8$ crystals with a thickness of ~ 10 nm (**e**) and 1.4 to 2.8 nm (the dashed lines serve as visual aids to discern the boundaries of 2D crystals) (**f**) on a single monolayer

WSe$_2$; **g,** An AFM image of an area of one unit cell thick Cr$_5$Te$_8$ crystals; the dashed line shows the boundary between monolayer WSe$_2$ and sapphire substrate.

Optical microscope images of Cr$_5$Te$_8$/WSe$_2$ heterostructures with relatively thick (~ 10 nm) and thin (1.4 to 2.8 nm) Cr$_5$Te$_8$ crystals are shown in Figure 1e and 1f, respectively. They were achieved by controlling the distance between the CrCl$_3$ precursor and the substrate. The atomically thin Cr$_5$Te$_8$ crystals in Figure 1f exhibit extremely weak contrast. To help discern these crystals from the substrate, the boundaries of these crystals are highlighted by the dashed lines (the same image with enhanced contrast is shown in Figure S2b, supporting information). A group of one unit cell thick, aligned Cr$_5$Te$_8$ crystals on WSe$_2$ are further shown by the atomic force microscopy (AFM) image in Figure 1g and Figure S2d.

The *a* lattice constants of freestanding WSe$_2$ and Cr$_5$Te$_8$ are 3.33 and 7.90 Å, respectively. Conventional vdW epitaxy mechanism implies that with ~ 16% lattice mismatch, defined as $(a_{Cr_5Te_8} - 2 * a_{WSe_2})/a_{Cr_5Te_8}$, the Cr$_5$Te$_8$ and WSe$_2$ would form incommensurate moiré superlattices[19]. Indeed, a moiré pattern can be clearly seen in the atomically resolved plane-view high angle annular dark field scanning transmission electron microscopy (HAADF-STEM) image in **Figure 2**a. However, the strictly periodic moiré pattern suggests that the Cr$_5$Te$_8$ and WSe$_2$ lattices are commensurate. The fast Fourier transform (FFT) pattern (Figure 2b) reveals three distinct sets of diffraction spots with six-fold symmetry, marked by yellow, red, and blue circles, respectively. The lattice spacing of the diffraction spots marked by yellow circles is 2.88 Å, which is consistent with the measured (100) lattice spacing of monolayer WSe$_2$. The diffraction spots marked with red circles show identical orientation with that of the WSe$_2$, with a lattice spacing of 3.36 Å in accordance with the Cr$_5$Te$_8$ (200) planes, which reduces by ~2% from the value of freestanding Cr$_5$Te$_8$[20]. The inner, red-circled diffraction spots with a larger periodicity (6.72 Å) representing those ordered, self-intercalated Cr atoms is a structural fingerprint of trigonal $Cr_5Te_8$. Interestingly, the innermost diffraction spots highlighted by blue circles indicate a lattice spacing of 11.6 Å, which belongs to neither $WSe_2$ nor $Cr_5Te_8$ alone, suggesting that it originates from the periodicity of the moiré superlattice of the $Cr_5Te_8/WSe_2$ heterostructure. The individual lattices of $Cr_5Te_8$ and $WSe_2$ are resolved by inverse FFT (iFFT) of the corresponding filtered diffraction spots (Figure S3, supporting information), showing hexagonal lattices of both $WSe_2$ and $Cr_5Te_8$ aligned in identical orientation, as seen in Figure 2c and 2d, respectively.

An atomic model of the $Cr_5Te_8/WSe_2$ heterostructure based on iterative refinements from HAADF-STEM image analysis and first principles calculations is shown in Figure 2e, where a $3 \times 3$ $Cr_5Te_8$ supercell is commensurate with the $7 \times 7$ $WSe_2$ supercell. To generate the larger periodicity with six-fold symmetry that reproduces the observed moiré diffraction pattern shown by the blue circles in Figure 2b, the model requires the number of interfacial Cr atoms to be reduced to 3 per supercell from 9 self-intercalated in the interior of $Cr_5Te_8$, suggesting interfacial reconstruction. In slight variations of the atomic model, either without interfacial Cr or with 9 interfacial Cr per supercell, the moiré superlattice diffraction pattern is noticeably missing since it is symmetry forbidden (Figure S5, supporting information). This further confirms the reconstructed interface with suitable Cr occupation is a necessary condition for the observed moiré diffraction pattern and the commensurate moiré superlattice. As can be seen from Figure 2f, the simulated diffraction pattern matches well with that of Figure 2b. The high-resolution STEM image reveals the perfectly commensurate moiré superlattice with a lattice constant of 23.3 Å, as shown in Figure 2g. The simulated STEM image (Figure 2h) based on the atomic model reproduces the moiré pattern observed. The surprising observation of the commensurate moiré superlattice is contradictory to the common knowledge of weak vdW interactions; instead, it suggests the presence of chemical bonding between $Cr_5Te_8$ and $WSe_2$.

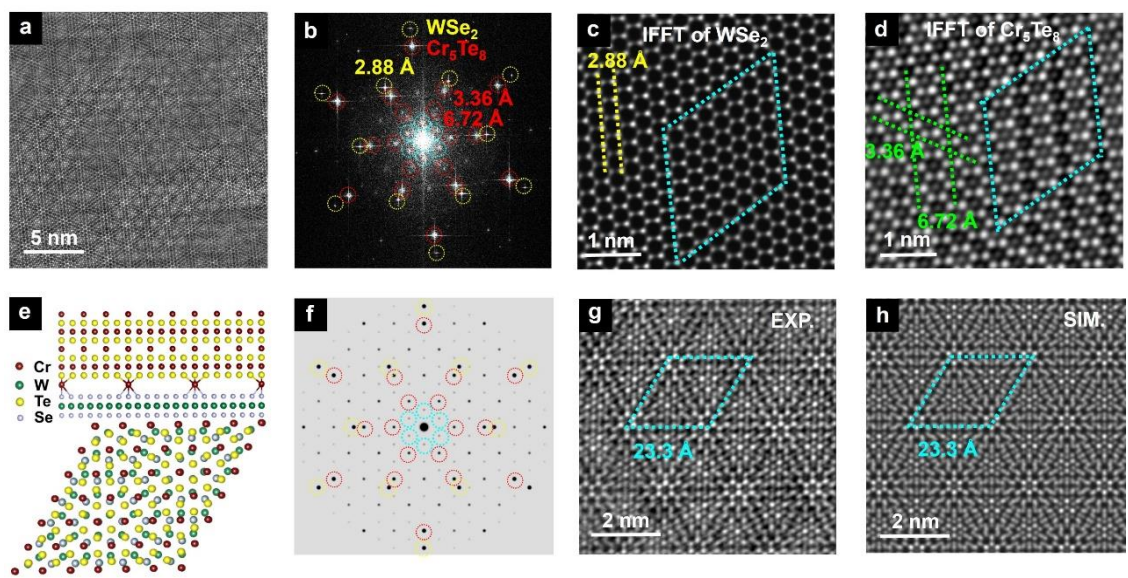

**Figure 2. HAADF-STEM analysis of the Cr₅Te₈/WSe₂ moiré superlattice. a,** Atomic-resolution HAADF-STEM image showing the moiré pattern of the $Cr_5Te_8$/$WSe_2$ heterostructure. **b,** FFT pattern obtained from (**a**). The diffraction spots marked by yellow circles: $WSe_2$ (100); outer red circles: $Cr_5Te_8$ (200); inner red circles: self-intercalated Cr atoms matched to trigonal $Cr_5Te_8$; blue circles: the moiré superlattice. **c, d,** IFFT images of identically oriented $WSe_2$ (**c**) and $Cr_5Te_8$ (**d**) lattices obtained from (**b**). **e,** Side view along (210) and top view of the atomic model of the $Cr_5Te_8$/$WSe_2$ superlattice. **f,** Simulated diffraction patterns obtained from the atomic model, matching that in (**b**). **g,** Experimental and **h,** simulated HAADF-STEM images showing identical moiré superlattice of $Cr_5Te_8$/$WSe_2$.

First principles calculations were carried out to understand the atomic structure, charge transfer, and chemical bonding at the $Cr_5Te_8$/$WSe_2$ interface. In bulk, the self-intercalated Cr atoms are coordinated with 6 Te atoms arranged on the corners of a triangular prism, as seen in Figure 2e. The interfacial Cr atoms have excess electrons due to the reduced coordination (losing 3 nearest Te neighbors comparing to the bulk). After forming the heterostructure, the crystal symmetry is further lowered with these interfacial Cr atoms coordinating to Se atoms on the $WSe_2$ side, forming a Te-Cr-Se Janus interface. **Figure 3**a compares the site-decomposed partial DOSs of interfacial Cr (left panel) and Se (right panel) sp states for isolated $Cr_5Te_8$ and $WSe_2$ monolayer (red curves) with those of $WSe_2$/$Cr_5Te_8$ heterostructure (black curves). As can be seen from the change of DOS of Cr in the left panel, the majority spin state peak near the Fermi level (the red peak) is suppressed. Accompanying the suppression, a band at ~ -7 to -4 eV emerges. The emergence of this low-lying Cr band is a result of hybridization with the Se sp states, whose energy is also lowered due to the hybridization, as can be seen in the right panel.

One can understand the above results based on a level repulsion picture in Figure 3b: the two states shown in red color are the interfacial Cr and Se sp states before $Cr_5Te_8$ and $WSe_2$ form the heterostructure. When forming the heterostructure, the Cr and Se sp states hybridize: the high-lying Cr sp state is pushed up in energy forming an anti-bonding state so its spectra weight near the Fermi level is suppressed, while the low-lying Se sp state is pushed down forming a bonding state in line with the result in the right panel in Figure 3a. As this is a hybridization, Cr also takes a significant share in the newly-formed low-lying state, as evidenced by the band at ~ -7 to -4 eV in the left panel in Figure 3a. Interestingly, after the hybridization, the antibonding state is pushed up to be above the $WSe_2$ conduction band edge, resulting in electron transfer from the interfacial Cr to $WSe_2$ conduction band to reduce system energy.

Figure 3c shows the differential (deformation) charge density $\Delta\rho$, which reveals a bonding-state charge accumulation in-between the interfacial Cr and Se atoms. Based on the Bader analysis[21] (see the Table S2), the total amount of charge accumulated is about 1 electron per supercell. The interfacial Cr-Se bond length of 2.88 Å is noticeably longer than the 2.56 Å for $Cr_2Se_3$. The interfacial binding energy of ~ 1 eV/Cr is an order of magnitude larger than a typical vdW binding but is only half of the usual Cr-Se covalent bond. These results point consistently to the formation of dative bonds at the interface[13], which originates from Coulomb attraction between anion lone pairs (*i.e.*, the doubly occupied non-bonding states of interfacial Se atoms) and the empty orbitals of the metal cations (*i.e.*, the nearly empty Cr sp states upon electron transfer to the $WSe_2$ conduction band) and is intermediate in strength between the vdW binding and covalent bond. The formation of the dative bonds weakens adjacent Cr-Te and W-Se bonds, *e.g.*, the W-Se bond length increases from 2.545 to 2.550 Å.

The formation of directional dative bonds is ultimately responsible for fixing the atomic registry and orientation of the $Cr_5Te_8$ 2D crystals on $WSe_2$ monolayer. It represents a new regime of thin film epitaxy that is distinctly different from either a conventional 3D epitaxy with strong covalent bond or a standard vdW epitaxy. We expect that dative epitaxy can be generally applicable to other covalent materials on vdW templates. The conditions for the formation of dative bonds at the interface are the presence of metal cations that can donate electrons and lone pairs that exist in many vdW materials such as transition metal chalcogenides.

Figure 3d plots the calculated bulk lattice parameters of $CrTe_x$ (represented by the superlattice parameter) as a function of the number of self-intercalated Cr atoms. It can be seen that the lattice parameter shrinks monotonically with decreasing the number of Cr atoms, *e.g.* from 23.8 Å for 9 Cr in $Cr_5Te_8$ to 23.3 Å for 3 Cr. Note 3 Cr is identical to the interfacial Cr number in the atomic model that reproduces the moiré diffraction pattern, and 23.3 Å is the measured moiré superlattice parameter that is exactly 7 times the lattice constant of monolayer $WSe_2$. It is remarkable that nature optimizes dative bond formation to remove the ~ 2% interfacial strain that would appear otherwise for the epitaxial growth.

In conventional vdW heterostructures with large lattice mismatch, either incommensurate superlattices with spatially varying moiré patterns[16] or local commensurate domain reconstruction were observed[17]. In conventional covalent heteroepitaxy, on the other hand, interfacial strain would lead to defects such as dislocations. For the $Cr_5Te_8/WSe_2$ system, however, nature optimizes the atomic structure with just the right number of dative bonds at the interface. This allows nearly strain-free commensurate moiré superlattices over the entire 2D heterostructure with minimum density of interfacial defects. As evidenced by the HAADF-STEM images taken at different spots of a single $Cr_5Te_8/WSe_2$ heterostructure (Figure S4, supporting information), identical, perfectly commensurate moiré patterns were observed across the heterostructure, suggesting that the $Cr_5Te_8$/$WSe_2$ is a monocrystalline moiré supercrystal. To our knowledge, such a monocrystalline moiré supercrystal has not been reported before and provides strong evidence for the proposed dative epitaxy mechanism.

The cross-section of a relatively thick (~ 7 nm) $Cr_5Te_8$ layer grown on $WSe_2$ were imaged by HAADF-STEM to reveal atomic details of the interface (atomically thin $Cr_5Te_8$/$WSe_2$ got oxidized during cross-sectional sample preparation). To reveal the position of Cr atoms that are much lighter than Te, integrated differential-phase contrast (iDPC) imaging technique were employed. As can be seen in Figure 3e, the cross-section of $Cr_5Te_8$ viewed along (100) direction matches well with the atomic model in Figure S1a. Atomic columns with substantially weaker contrast inside the vdW-like gap can be seen from the zoomed-in view in Figure 3f, which is attributed to the reduced number of interfacial Cr atoms, consistent with the atomic model matching diffraction and theoretical predictions. The local valence states of Cr atoms across the interface were mapped using the integrated intensity ratio of the electron energy loss spectroscopy (EELS) $L_3$ and $L_2$ excitation peaks (the so-called "white line ratio")[22]. As shown in Figure 3g, the Cr $L_3$/$L_2$ ratio increases towards the interface and becomes substantially larger than the value in the $Cr_5Te_8$ interior, indicating a lower Cr valence state near the interface (individual EELS spectrum at the bulk and interface is provided in Figure S6, supporting information). This is because after dative bond formation and electron donation, these interfacial Cr atoms still possess excessive charge due to lower coordination.

The predicted weakening of W-Se bonds in $WSe_2$ was further investigated by Raman spectroscopy. As seen from the bottom panel in Figure 3h, a strong peak at ~ 250 cm$^{-1}$ and a weak shoulder at ~ 260 cm$^{-1}$ are observed for monolayer $WSe_2$, which can be attributed to the degenerate out-of-plane $A_{1g}$ and in-plane $E_{2g}^1$ phonon modes of $WSe_2$, and a second-order

Raman mode due to LA phonons at the M point in the Brillouin zone labeled as 2LA(M), respectively[23]. These modes are also observed for the $Cr_5Te_8/WSe_2$ moiré superlattice. However, both peaks showed a small but measurable red shift. The average Raman peak positions measured at 9 different spots each for $WSe_2$ and $Cr_5Te_8/WSe_2$ are shown in Table 1 (all spectra and fittings are shown in Figure S7a, b and Table S1). A red shift ($\Delta\nu$) of 1.3 cm$^{-1}$ for the $E_{2g}^1/A_{1g}$ mode and 2.4 cm$^{-1}$ for the 2LA(M) mode were observed, confirming the predicted W-Se bond softening due to dative bond formation.

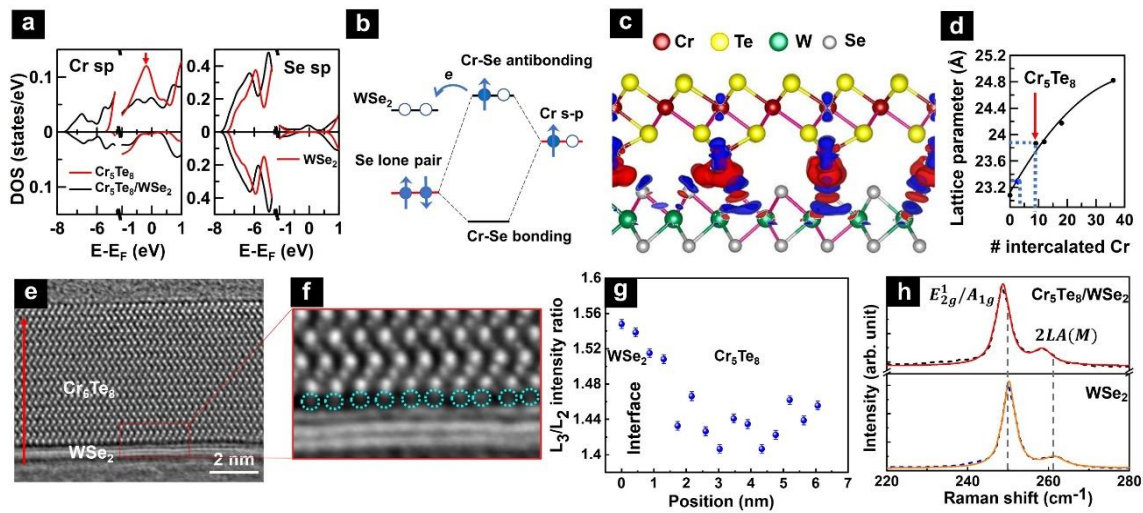

**Figure 3. Atomic and electronic structure of the $Cr_5Te_8/WSe_2$ interface. a,** Site-decomposed partial DOSs of interfacial Cr (left) and Se (right) sp states for indvidual $Cr_5Te_8$, $WSe_2$ (red curves) and $WSe_2/Cr_5Te_8$ (black curves). **b,** A schematic diagram illustrating the dative bond formation process. **c,** c-axis projected differential charge density $\Delta\rho$ profile along the Te-Cr-Se-W direction. **d,** Calculated lattice spacing of $CrTe_x$ as a function of self-intercalated Cr number. **e,** Cross-sectional iDPC image of the $Cr_5Te_8/WSe_2$ heterostructure. The red arrow in (**e**) indicates the EELS line scan direction; the red box is the area shown in (**f**). **f,** A magnified image in (**e**). The circles mark the atomic columns with weak contrast, which are attributed to interfacial Cr atoms. **g,** The integrated intensity ratio between Cr $L_3$ and $L_2$ edges measured by EELS as a function of position from $WSe_2$ to $Cr_5Te_8$ along the red arrow direction in (**e**). Error bars represent statistical uncertainty of the mean value. **h,** Raman spectra of monolayer $WSe_2$ and $Cr_5Te_8/WSe_2$ heterostructure. Black dashed lines: data; red and orange lines: fittings using Lorentzian function.

| Table 1. Raman peak positions of monolayer WSe₂ and Cr₅Te₈/WSe₂. | | | | | |
|---|---|---|---|---|---|
| WSe₂ $E^1_{2g}/A_{1g}$ (cm⁻¹) | Cr₅Te₈/WSe₂ $E^1_{2g}/A_{1g}$ (cm⁻¹) | $\Delta v$ $E^1_{2g}/A_{1g}$ (cm⁻¹) | WSe₂ $2LA(M)$ (cm⁻¹) | Cr₅Te₈/WSe₂ $2LA(M)$ (cm⁻¹) | $\Delta v$ $2LA(M)$ (cm⁻¹) |
| 250.21±0.04 | 248.93±0.12 | 1.3 | 261.63±0.08 | 259.24±0.18 | 2.4 |

Dative epitaxy enables nearly strain-free epitaxial growth of discrete monocrystalline 2D $Cr_5Te_8$ crystals on a single monolayer $WSe_2$, which should lead to extremely low density of interfacial defects. The out-of-plane magnetic hysteresis of single 2D $Cr_5Te_8$ crystals were measured by reflective magnetic circular dichroism (RMCD), which is used to infer the crystallinity of $Cr_5Te_8$. Shown in **Figure 4**a-c are three representative 2D $Cr_5Te_8$ crystals with thicknesses of 8.4 nm (6 unit cells), 4.5 nm (3 unit cells) and 2.6 nm (2 unit cells), respectively. The magnetic hysteresis loops for the three crystals measured at 5 K are shown in Figure 4d-f. All three samples exhibit square hysteresis loops, with sharp transitions at the coercive fields ($H_C$), and unity remanence at zero field. $H_C$ are 0.66, 0.36, and 0.74 T for 6, 3 and 2 unit cell thick crystals, respectively, which are noticeably smaller than the expected anisotropy field[24], suggesting that the magnetization reversal proceeds by nucleation (*e.g.* at a sharp corner) followed by domain wall motion. The nearly perfect square hysteresis suggests nearly absence of domain wall pinning by defects, and thus once a magnetic domain is nucleated, the domain wall can propagate freely. On the other hand, the reported magnetic hysteresis of $Cr_xTe_{1-x}$ nanoplates on covalent substrates show skewed loops with a broad switching field distribution[25]. This might imply the presence of interfacial pinning sites. This comparison suggests that 2D $Cr_5Te_8$ crystals obtained by dative epitaxy does possess superior crystal quality and magnetic properties. For one unit cell thick $Cr_5Te_8$ crystals, however, no RMCD signal can be detected. The lack of magnetic signal may be due to surface oxidation.

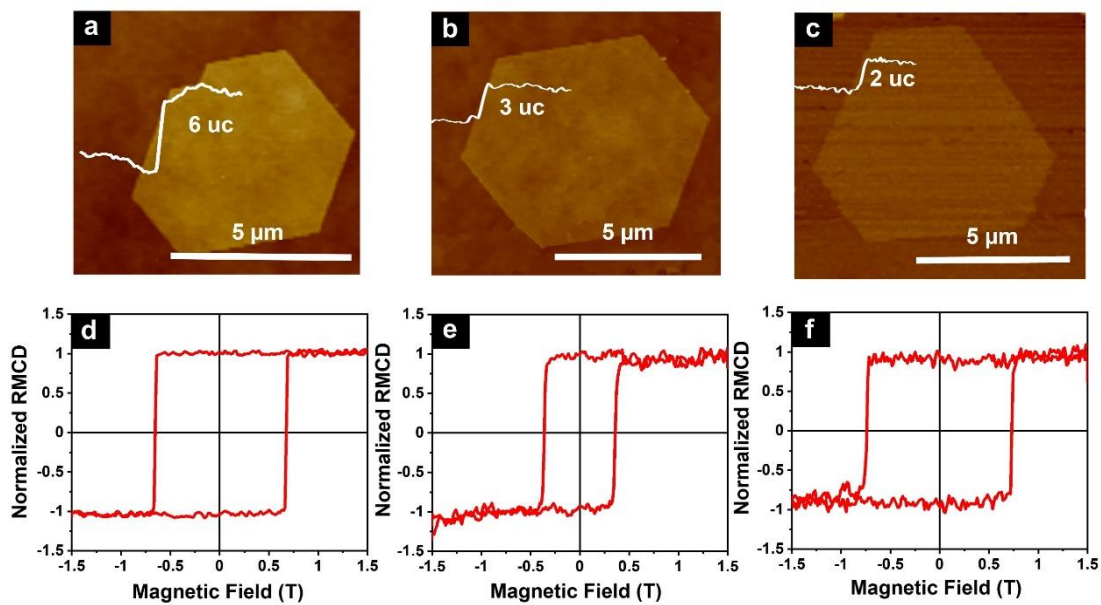

**Figure 4. RMCD measurements of Cr$_5$Te$_8$ single flakes grown on WSe$_2$. a, b, c,** AFM images of Cr$_5$Te$_8$ crystals grown on WSe$_2$ and **d, e, f,** the corresponding RMCD hysteresis loops measured at 5 K.

## 3. Conclusion

In conclusion, Cr$_5$Te$_8$/WSe$_2$ moiré supercrystal has been achieved by dative epitaxy of covalent 2D Cr$_5$Te$_8$ crystal on monolayer WSe$_2$. The Cr-Se dative bond formation at the interface drives the epitaxial growth of highly aligned 2D Cr$_5$Te$_8$ crystals. The dative epitaxy results in perfectly commensurate, monocrystalline moiré supercrystals, which are distinctly different from conventional moiré superlattices. The high crystal quality of the 2D Cr$_5$Te$_8$ crystals is further confirmed by square magnetic hysteresis loops nearly free of defect pinning sites. The dative epitaxy is likely not limited to the present material systems and synthesis method, but should be applicable to a wide range of covalent materials on vdW templates, using a variety of thin film deposition techniques. This newly realized paradigm of thin film epitaxy not only is attractive for a plethora of industrial applications, but also offers opportunities to explore emergent phenomena in previously unattainable heterostructures.

## 4. Methods

**Synthesis of $Cr_5Te_8$/$WSe_2$ heterostructures:** $Cr_5Te_8$/$WSe_2$ heterostructures were synthesized through a two-step CVD process in a two-zone tube furnace with a 2" diameter. A schematic of the experimental setup and the heating profiles of the synthesis were shown in Figure S1b, c. In the first step, $WSe_2$ monolayer was grown on sapphire or $SiO_2$/Si substrate. In a typical synthesis, 200 mg Se powder was placed in the first heating zone upstream of the furnace, which was kept at 400 °C during the growth. 5 mg $WO_3$ was mixed with 0.5 mg NaCl and loaded in the second heating zone downstream. Sapphire substrates were placed close to the $WO_3$ powder, while $SiO_2$/Si substrates were placed face down directly above the $WO_3$ powder. The second heating zone was heated with a ramping rate of 20 °C/min to the growth temperature of 820 °C and held at that temperature for 20 min before cooling down. During the growth process, the flow rate of 2% $H_2$/$N_2$ was kept at 80 standard cubic centimeters (sccm), and the growth was at ambient pressure. In the second step, the as-grown $WSe_2$ were used as the template for the epitaxial growth of 2D $Cr_5Te_8$ crystals to obtain $Cr_5Te_8$/$WSe_2$ heterostructures. 40 mg Te powder was placed in the first heating zone and ramped to 540 °C at a rate of 13 °C/min, and kept at 540 °C for 10 min. 1.2 mg $CrCl_3$ powder was placed in the second heating zone, and heated to 600 °C at a rate of 15 °C/min. The growth time was fixed at 10 min. A gas mixture containing 90% Ar and 10% $H_2$ with a flow rate of 100 sccm was used to carry the precursor vapor species to the substrate. Once the reaction ended, the furnace was cooled down naturally to room temperature.

The CVD growth of 2D $Cr_5Te_8$ crystals on $WSe_2$ template is dictated by monomer adsorption, desorption, and surface diffusion, as seen in Figure 1a. The weak bonding between the monomer and the $WSe_2$ vdW template leads to low barriers for surface diffusion[26]. The monomers aggregate to form nuclei. Once a nucleus reaches a critical size,[26a] the growth proceeds primarily by surface diffusion of monomers and their attachment to the edges of the nucleated 2D islands[26b]. For $Cr_5Te_8$, the intralayer covalent bonding is substantially stronger than the interlayer bonding due to the presence of ordered vacancies. Such bonding character results in stronger adsorption energy of the monomers at the edges than that on the top surfaces. Together with ease of diffusion of monomers on the surface of $WSe_2$, atomically thin $Cr_5Te_8$ crystals can be achieved[26b].

The $Cr_5Te_8$/$WSe_2$ heterostructures with relatively thick (~ 10 nm) and thin (1.4 to 2.8 nm) $Cr_5Te_8$ crystals were achieved by controlling the distance between the $CrCl_3$ precursor and the substrate, being ~ 0.2 mm for the samples shown in Figure 1e and ~ 2 mm for the ones in Figure 1f. Due to the high melting point (1,150 °C) of $CrCl_3$, a steep vapor concentration gradient was established at the growth temperature of 600 °C. At a large precursor-substrate distance, the vapor concentration can be kept below the threshold of new nucleation on top of existing 2D layers, resulting in atomically thin $Cr_5Te_8$ crystals.

Interestingly, the 2D $Cr_5Te_8$ crystals deposit highly selectively on $WSe_2$ only, leaving sporadic nanocrystals on sapphire, as seen in Figure 1d. We attribute the absence of 2D crystal growth on these substrates to the covalent bonding and thus large surface diffusion barrier for monomers. This leads to 3D growth of nanoparticles which eventually dewet due to surface tension.

**Film transfer**: The as-grown $Cr_5Te_8$/$WSe_2$ heterostructures were transferred onto TEM grids by dry transfer in a glovebox with a nitrogen atmosphere. The $Cr_5Te_8$/$WSe_2$ heterostructures on $SiO_2$/Si substrate were first covered by polymethylmethacrylate (PMMA)[27]. After baking at 80 °C for 5 min, the PMMA film with $Cr_5Te_8$/$WSe_2$ heterostructures was peeled off from the $SiO_2$/Si substrate and then transferred to a TEM grid in a home-built alignment stage integrated with an optical microscope, followed by 5 min baking at 80 °C. The PMMA was removed by immersing the sample in acetone for 30 min.

**Cross-sectional STEM sample preparation:** The as-grown $Cr_5Te_8$/$WSe_2$ heterostructure was exposed to a nitrogen atmosphere and subsequently covered by graphite through a routine dry-transfer method in the glovebox to protect the surface from being oxidized. The cross-section STEM sample was prepared by using Focused Ion Beam (FIB) milling. It was thinned down to 70 nm thick at an accelerating voltage of 30 kV with a decreasing current from 0.79 nA to 80 pA, followed by a fine polish at an accelerating voltage of 2 kV with a small current of 21 pA to remove the amorphous layer.

**HAADF-STEM characterization:** The atomically resolved HAADF-STEM images were carried out on an aberration-corrected scanning transmission electron microscope (FEI Tian Themis 60-300kV, operate at 300 kV). This TEM is equipped with a DCOR aberration corrector and a high-brightness field emission gun (X-FEG) with monochromator. The inner and outer collection angles for the STEM images ($\beta 1$ and $\beta 2$) were 38 and 200 mrad, respectively, with a semi convergence angle of 30 mrad.

**Cross-sectional STEM imaging:** The cross-sectional HAADF-STEM reveals a clear vdW-like gap between the $Cr_5Te_8$ and $WSe_2$ layers as shown in Figure S4h, but Cr sites are nearly invisible due to strong scattering of Te which has a much larger atomic weight. To reveal the position of Cr, we adopted integrated differential-phase contrast (iDPC) imaging, which measures the projected electrostatic potential instead of the integrated scattering signal of the atomic column.

**STEM-EELS characterization:** EELS were acquired in the STEM mode and collected by setting the energy resolution to 1 eV at full width at half maximum (FWHM) of the zero-loss peak. The dispersion used is 0.5 eV/channel. EELS are acquired in the dual EELS mode to eliminate any systematic error due to the drift of the zero-loss peak.

**STEM-EDS characterization**: EDS was acquired in the STEM mode with a ChemiSTEM technology (X-FEG and SuperX EDS with four windowless silicon drift detectors) operated at 300 kV.

**Raman and photoluminescence spectra** were measured using a confocal Renishaw inVia

Raman microscope equipped with a 514 nm laser. A 50× objective lens was used to focus the excitation lasers onto the sample and collect the emitted signals.

**X-ray diffraction (XRD) spectrum**: A Rigaku Ultima IV XRD system with an operational X-ray tube power of 1.76 kW (40 kV, 44 mA) and Cu target source was used. The XRD measurements were performed under theta/2 theta scanning mode and continuous scanning type with a step size of 0.02.

**X-ray photoelectron spectroscopy (XPS) spectrum:** XPS was conducted on a PHI 5000 Versaprobe system using Al Kα X-ray radiation for excitation.

**DFT-based ab-initio calculations** were performed by using the Vienna *ab initio* Simulation Package (VASP) package. We used the Perdew–Burke–Ernzerhof (PBE) form of the exchange correlation functional. Slab calculations were performed using supercell approach with a vacuum layer of ~15 Å (to remove interaction between periodically repeated layers). The supercell was constructed using the observed moiré superlattice. The in-plane lattice constant of Moiré superlattice was set at 23.3 Å. Plane-wave cut-off energy of 400 eV, and 4 × 4 × 1 Monkhorst-Pack k-point mesh. The atomic positions were optimized by the conjugate gradient method to have all forces less that $10^{-2}$ eV/Å. Spin-orbit was added after the relaxation accuracy was achieved. Zero damping DFT-D3 method of Grimme models Van der Waals interactions.

**RMCD measurements**: The samples for RMCD measurements were capped with 2 nm Al by sputter deposition to prevent oxidization. The RMCD is defined as $(I_{\sigma+} - I_{\sigma-})/(I_{\sigma+} - I_{\sigma-})$, where the $I_{\sigma\pm}$ are the intensities of the reflected right and left circularly polarized light. RMCD measurements were performed with the sample mounted on a custom microscope/nano positioner probe that was loaded into the variable-temperature helium insert of a 7 T magneto-optical cryostat (Oxford Instruments Spectramag). Light from a 633 nm HeNe laser was linearly polarized, and then modulated between left- and right-circular polarization at 50 kHz using a photoelastic modulator, before being focused to a 1-micron diameter spot on the sample. Light reflected from the sample was detected by an avalanche photodiode, and the normalized difference between the two polarizations was measured using a lock-in amplifier.

**XMCD measurements:** XMCD measurements were performed at beamline 11.0.1 of the Advanced Light Source at the Lawrence Berkeley National Laboratory.

**Supporting Information**

Supporting Information is available from the Wiley Online Library or from the author.

**Acknowledgements**

H.Z., M.B., C.H., and A.M. acknowledge support from US National Science Foundation (ECCS-2042085, MRI-1229208, MRI-1726303, CBET-1510121), and University at Buffalo VPRED seed grant. Y.H. and M.B. acknowledge support from National Key R&D Program of China (2017YFA0206301), National Natural Science Foundation of China (52027801, 51631001, 52101280), China-German Collaboration Project (M-0199), and China Postdoctoral Science Foundation (2020M670042). J.L. and L.Z. acknowledge the support from the National Natural Science Foundation of China (Grant No.11974156), Guangdong International Science Collaboration Project (Grant No. 2019A050510001), the Science, Technology and Innovation Commission of Shenzhen Municipality (No. ZDSYS20190902092905285), and also the assistance of SUSTech Core Research Facilities. S.C. and J.C. acknowledge the support from National Science Foundation (DMR-1644779), the State of Florida, and the U.S. Department of Energy. X.M.C. and X.W. acknowledge the support from US National Science Foundation (DMR-1708790). S.Z. acknowledges the support from NSF ECCS-2042126. R.S. acknowledges the support from NU Collaborative Research and NSF-DMREF (1729288). This research used resources of the Advanced Light Source, which is a DOE Office of Science User Facility under contract no. DE-AC02-05CH11231.

**Author Contributions**

H.Z. and Y. H. conceived the project. H.Z., Y.H., and J.L. supervised the project. M.B., C.H., and A.M. prepared $Cr_5Te_8/WSe_2$, $Cr_5Te_8$, and $WSe_2$ samples. L.Z. and Q.Y. transferred samples and L.Z. performed HAADF-STEM, EDS, and EELS characterizations. J.C. and S.C. performed RMCD measurements. X.W. and R.V.C. performed XMCD measurements. R.S. performed the first-principal calculations. M.B. performed XRD, Raman, PL, and AFM measurements. Y.C.L. performed XPS measurements. H.Z., M.B., J.L., S.Z., R.S. and Y.H. wrote the manuscript. All authors discussed the results and commented on the manuscript.

**Conflict of Interest**

The authors declare no conflict of interest.

**Data Availability Statement**

The data that support the plots within this paper and other finding of this study are available from the corresponding author on reasonable request.

Dative epitaxy represents the Godilock's principle of epitaxy: it takes advantage of dative bonding for fixing the atomic registry and crystal orientation, while ensuring the full flexibility of vdW epitaxy. The globally commensurate $Cr_5Te_8/WSe_2$ moiré supercrystal is distinctly different from conventional incommensurate moiré superlattices or local commensurate domains.

M. Bian, L. Zhu, X. Wang, J. Choi, R. V. Chopdekar, S. Wei, L. Wu, C. Huai, A. Marga, Q. Yang, Y. C. Li, F. Yao, T. Yu, S. A. Crooker, X. M. Cheng, R. F. Sabirianov, S. Zhang, J. Lin*, Y. Hou* & H. Zeng*

**Dative epitaxy of commensurate monocrystalline covalent-van der Waals moiré supercrystal**

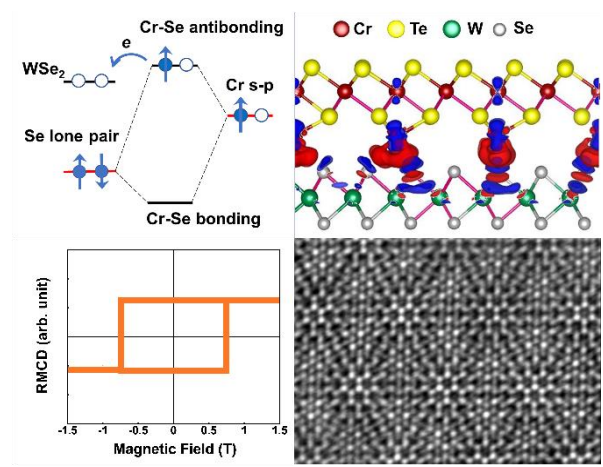

# Supporting Information

## Dative epitaxy of commensurate monocrystalline covalent-van der Waals moiré supercrystal

*Mengying Bian[1,2]†, Liang Zhu[3]†, Xiao Wang[4], Junho Choi[5], Rajesh V. Chopdekar[6], Sichen Wei[7], Lishu Wu[8], Chang Huai[2], Austin Marga[2], Qishuo Yang[3], Yuguang C. Li[9], Fei Yao[7], Ting Yu[8], Scott A. Crooker[5], Xuemei M Cheng[4], Renat F. Sabirianov[10], Shengbai Zhang[11], Junhao Lin[3]\*, Yanglong Hou[1]\* & Hao Zeng[2]\**

[1]Beijing Key Laboratory for Magnetoelectric Materials and Devices, Beijing Innovation Center for Engineering Science and Advanced Technology, School of Materials Science and Engineering, Peking University, Beijing, China

[2]Department of Physics, University at Buffalo, State University of New York, Buffalo, NY, USA

[3]Department of Physics and Shenzhen Key Laboratory of Advanced Quantum Functional Materials and Devices, Southern University of Science and Technology, Shenzhen, China

[4]Physics Department, Bryn Mawr College, Bryn Mawr, PA, USA

[5]National High Magnetic Field Laboratory, Los Alamos National Laboratory, Los Alamos, NM, USA

[6]Advanced Light Source, Lawrence Berkeley National Laboratory, Berkeley, CA, USA

[7]Department of Materials Design and Innovation, University at Buffalo, The State University of New York, Buffalo, NY, USA

[8]Division of Physics & Applied Physics, School of Physical and Mathematical Sciences,

Nanyang Technological University, Singapore

[9]Department of Chemistry, University at Buffalo, The State University of New York, Buffalo, NY, USA

[10]Department of Physics, University of Nebraska-Omaha, Omaha, NE, USA

[11]Department of Physics, Rensselaer Polytechnic Institute, Troy, NY, USA

†These authors contributed equally: Mengying Bian, Liang Zhu

*Corresponding author. e-mail: haozeng@buffalo.edu; hou@pku.edu.cn; linjh@sustech.edu.cn

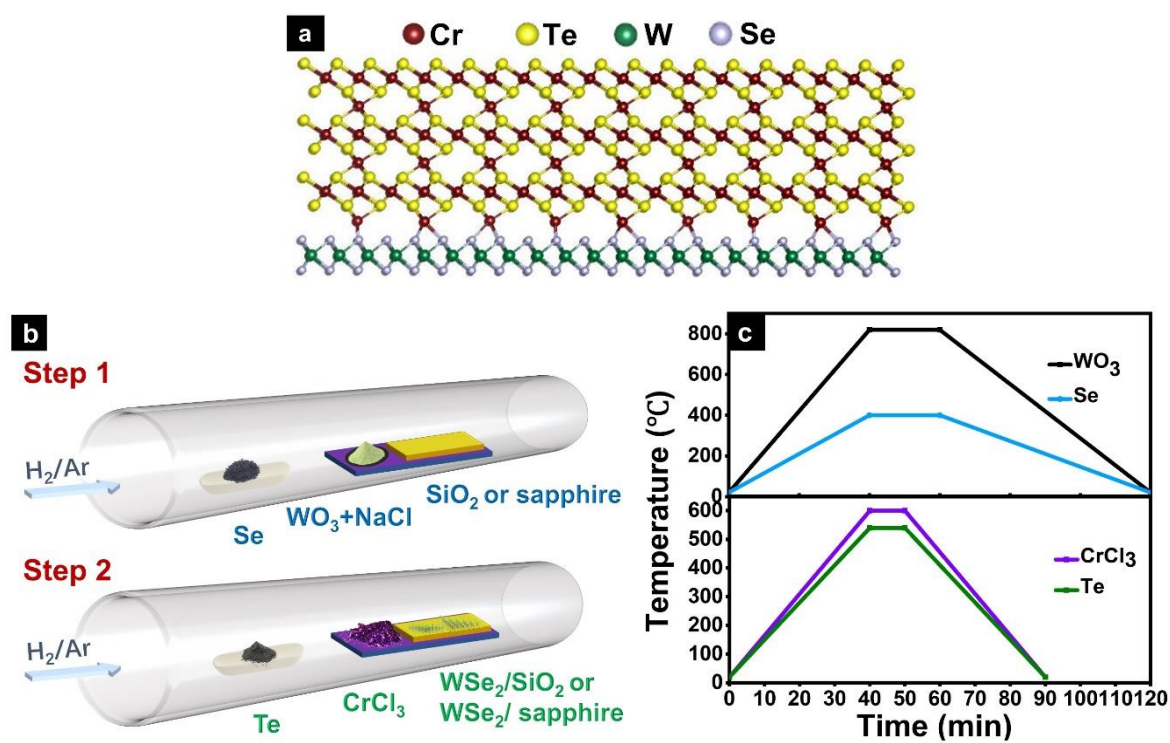

**Figure S1. CVD growth process of Cr$_5$Te$_8$/WSe$_2$ heterostructures and atomic model of Cr$_5$Te$_8$. a,** An atomic model of Cr$_5$Te$_8$/WSe$_2$ superlattice, as viewed along the (100) axis. **b,** A schematic diagram of the CVD set up for the growth of Cr$_5$Te$_8$/WSe$_2$ heterostructures. **c,** The heating profiles of the two zones of the two-step CVD growth process.

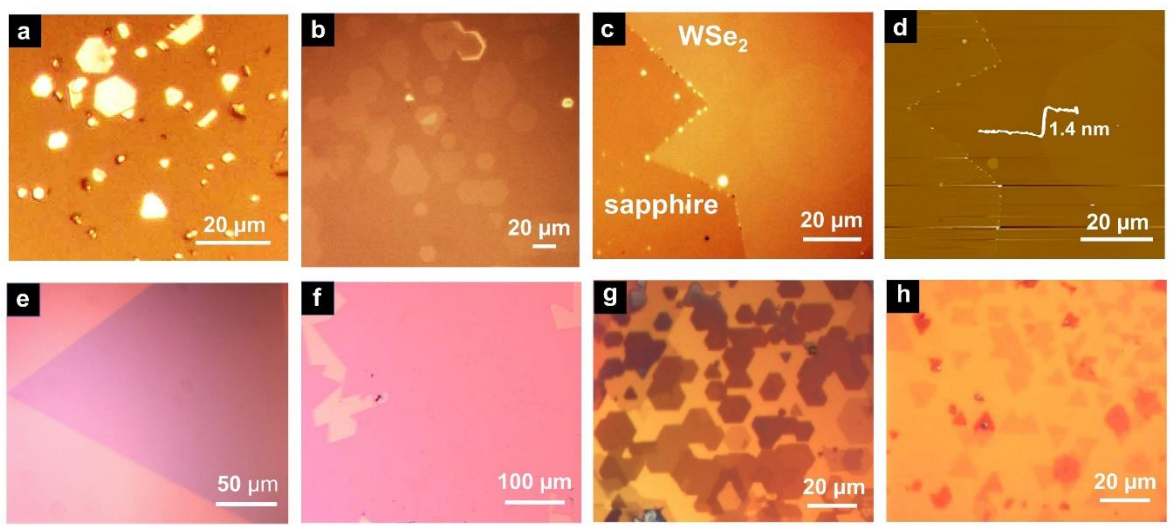

**Figure S2. Optical microscope images of Cr₅Te₈, WSe₂, Cr₅Te₈/WSe₂ heterostructures and atomic force microscope images.** Optical microscope images of **a,** random oriented $Cr_5Te_8$ crystals on sapphire substrate, and **b,** thin $Cr_5Te_8$ crystals of 1.4 to 2.8 nm on WSe₂ grown on sapphire, which is 2 mm away from the source (Same as Figure 1f, but with artificially enhanced contrast). **c,** An optical microscope image and **d,** the corresponding AFM image of a 2D $Cr_5Te_8$ crystal with a thickness of 1.4 nm (single unit cell) and the lateral size of ~46 µm grown on WSe₂ on sapphire. **e,** A triangular-shaped monolayer WSe₂ crystal on sapphire substrate. **f,** Part of a monolayer WSe₂ with a lateral dimension of ~ 1 mm. **g,** Thick (~10 nm) and dense $Cr_5Te_8$ 2D crystals grown on WSe₂ on SiO₂, which is 0.2 mm away from the source and **h,** thin $Cr_5Te_8$ crystals of one- to two unit cells thickness (1.4 to 2.8 nm) on WSe₂ on SiO₂, which is 2 mm away from the source.

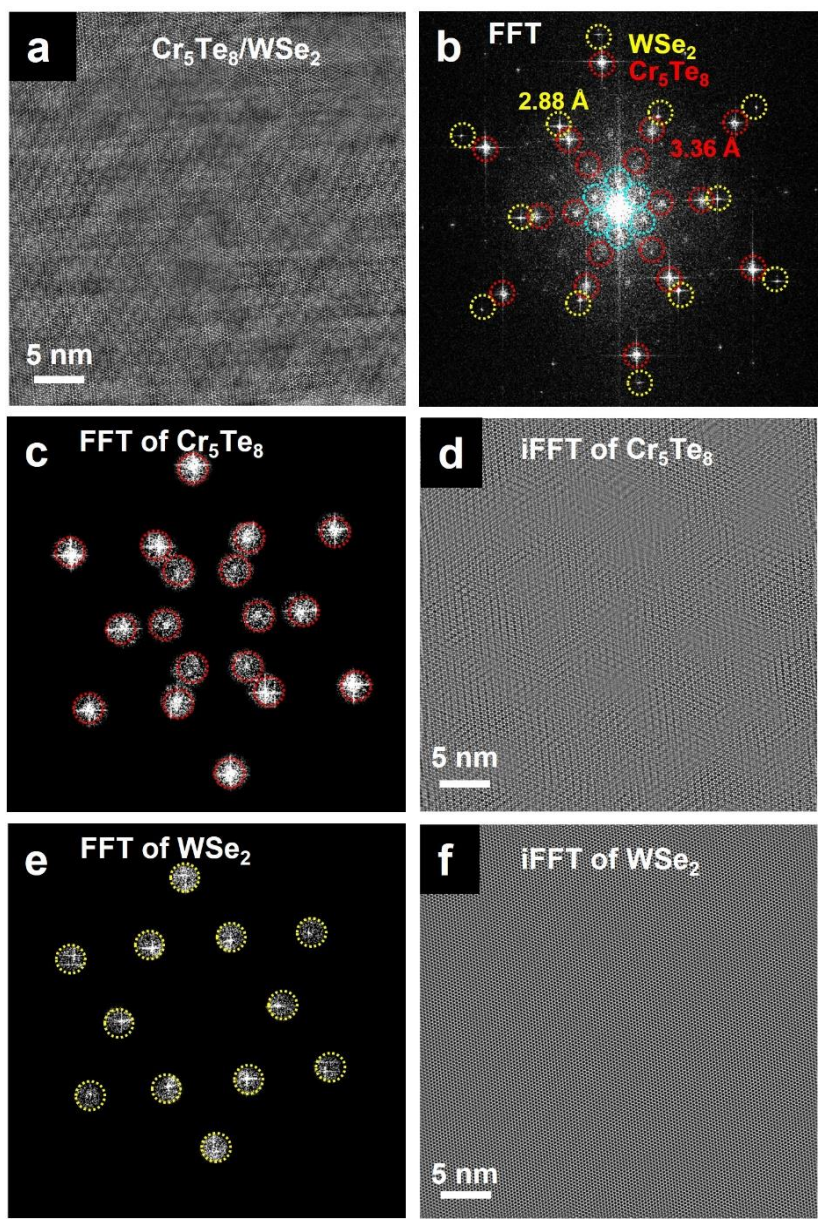

**Figure S3. A larger area atomic-resolution HAADF-STEM image of Cr₅Te₈/WSe₂ moiré superlattice**. **a, b,** A HAADF-STEM image of $Cr_5Te_8/WSe_2$ moiré superlattice and the corresponding FFT pattern obtained from (**a**). **c, d,** Selected diffraction point from $Cr_5Te_8$ and the corresponding iFFT image of $Cr_5Te_8$. **e, f,** Selected diffraction point from $WSe_2$ and the corresponding iFFT image of $WSe_2$.

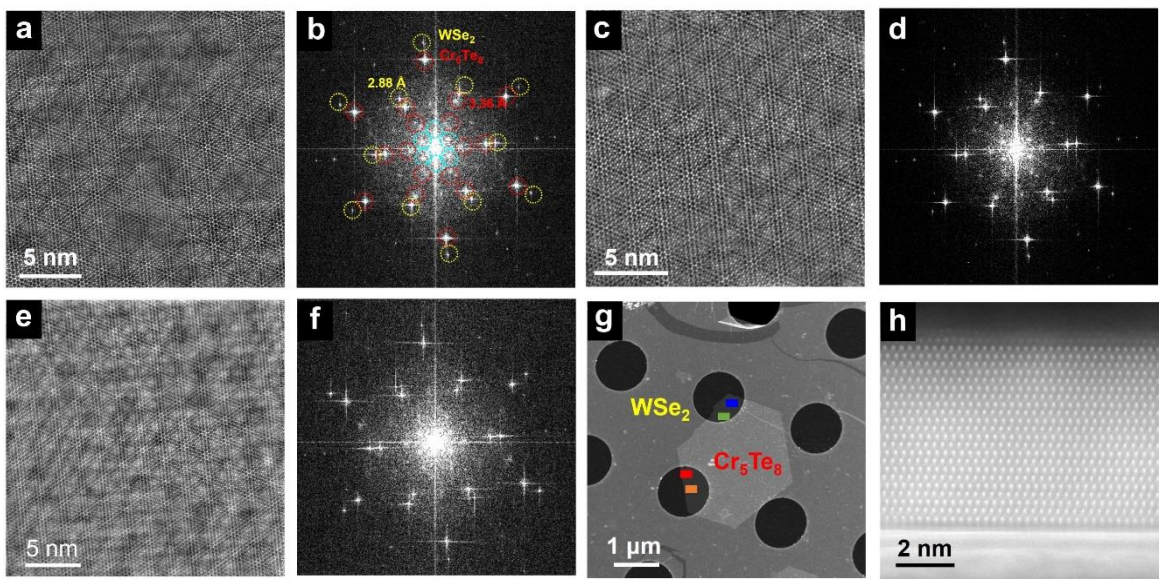

**Figure S4. HAADF-STEM images of Cr$_5$Te$_8$/WSe$_2$ moiré superlattices measured at different locations. a, c, e,** Atomic-resolution HAADF-STEM images showing the moiré pattern of the Cr$_5$Te$_8$/WSe$_2$ heterostructure from the blue, green, orange region in **g**. The atomic-resolution HAADF-STEM image of the region shown in red is shown in Figure 2. **b, d, f,** The corresponding FFT images of **a, c, e,** respectively. The moiré superlattice diffraction is marked by blue circles. **g,** Low resolution HAADF-STEM image of a single 2D Cr$_5$Te$_8$/WSe$_2$ heterostructure. **h,** A cross-sectional HAADF-STEM image of Cr$_5$Te$_8$/WSe$_2$ heterostructure.

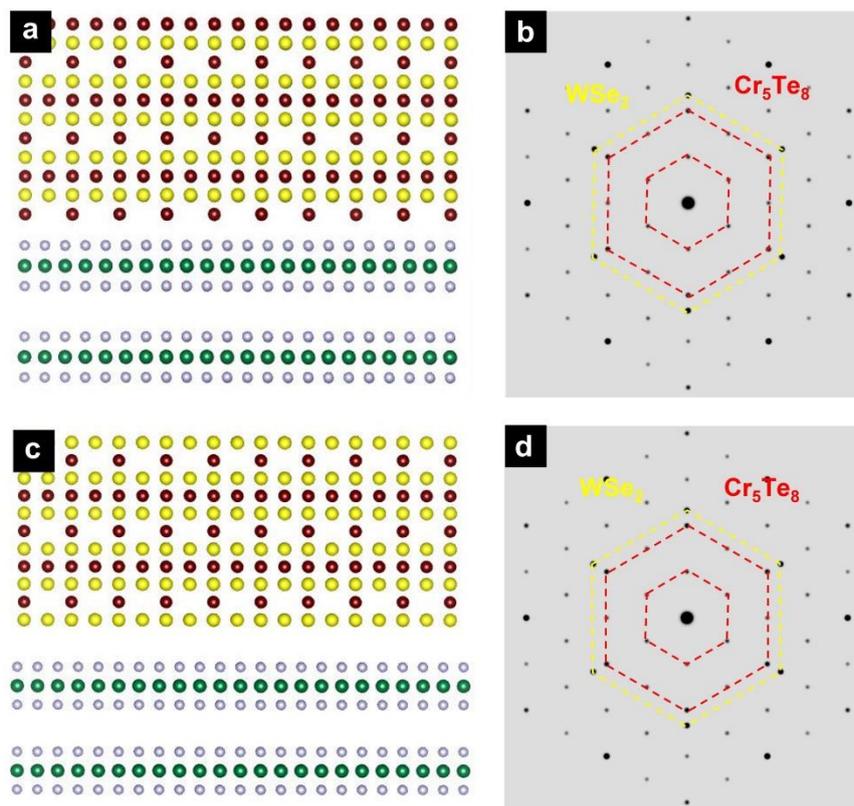

**Figure S5. Hypothetical atomic structures and simulated electron diffraction patterns of Cr₅Te₈/WSe₂ heterostructures**. **a,** Cross-sectional view of the atomic model with 9 interfacial Cr atoms per supercell, identical to the number of self-intercalated Cr in between CrTe₂ layers of Cr₅Te₈. **b.** The corresponding simulated electron diffraction of Cr₅Te₈/WSe₂ heterostructure (**a**). **c,** Cross-sectional view of the atomic model with Te-terminated interface. **d.** The corresponding simulated electron diffraction of Cr₅Te₈/WSe₂ heterostructure (**c**). The lattice periodicity belonging to the commensurate moiré superlattice is absent in both diffraction patterns.

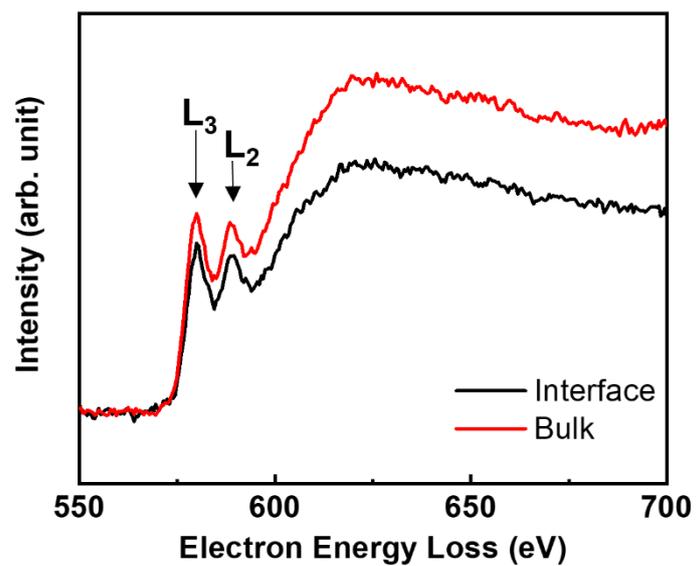

**Figure S6. EELS-STEM of Cr-L$_{2,3}$ and Te-M edge from the bulk and interface.** The integrated L$_3$/L$_2$ ratio is larger at the interface, suggesting lower valence state of interfacial Cr atoms.

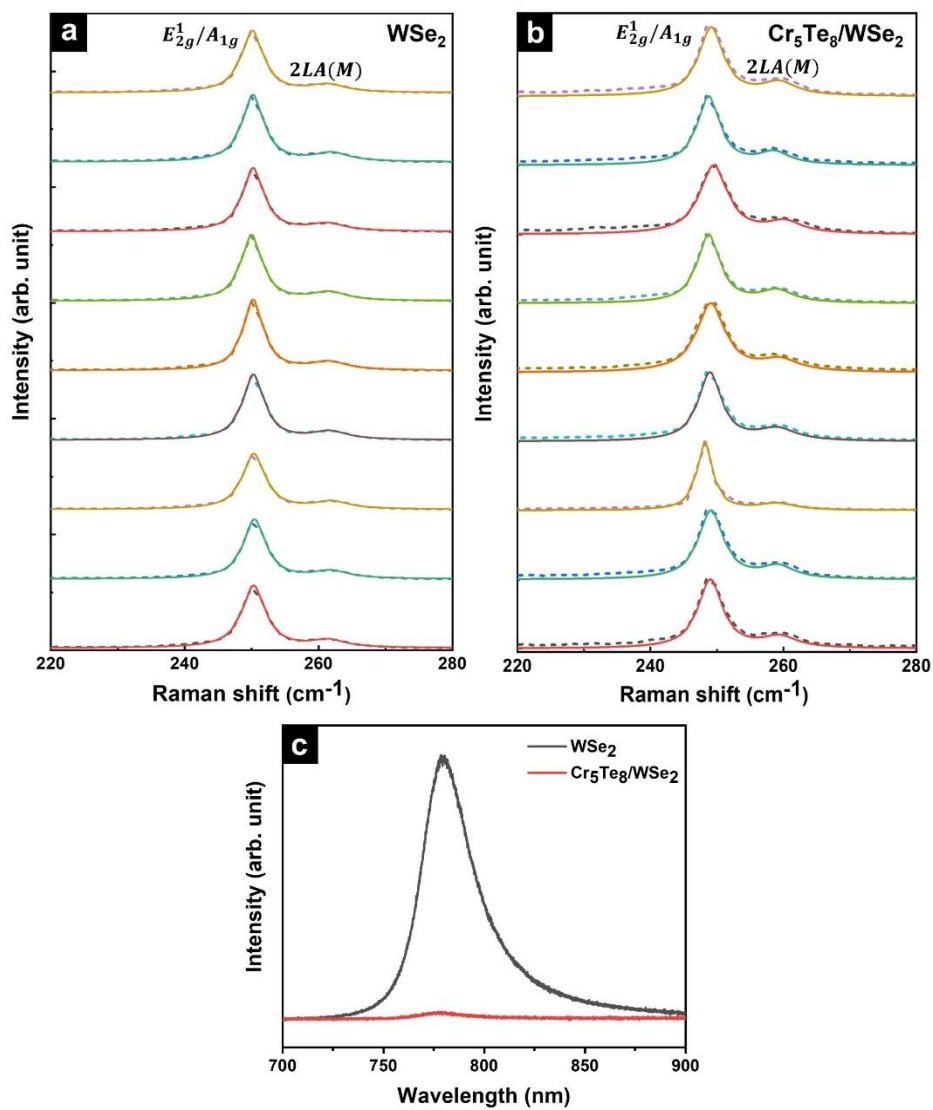

**Figure S7. Raman spectra and PL spectra of monolayer WSe₂ and Cr₅Te₈/WSe₂ heterostructure.** Raman spectra of representative **a,** monolayer WSe₂ and **b,** Cr₅Te₈/WSe₂ heterostructures measured at room temperature excited at 514 nm (dashed lines). The solid lines are cumulative fitting spectra using Lorentzian functions. **c,** PL spectra of monolayer WSe₂ and Cr₅Te₈/WSe₂ heterostructure grown on SiO₂/Si substrates were measured at room temperature. The significant decrease of PL intensity in the heterostructure can be attributed to the charge transfer of photo-induced carriers at the interface.

**Table S1. The representative Raman peak position of monolayer WSe₂ and Cr₅Te₈/WSe₂ heterostructure**

| WSe$_2$ $E_{2g}^1/A_{1g}$ （cm$^{-1}$） | WSe$_2$ $2LA(M)$ （cm$^{-1}$） | Cr$_5$Te$_8$/WSe$_2$ $E_{2g}^1/A_{1g}$ （cm$^{-1}$） | Cr$_5$Te$_8$/WSe$_2$ $2LA(M)$ （cm$^{-1}$） |
|---|---|---|---|
| 250.28 | 261.47 | 248.99 | 259.44 |
| 250.42 | 261.85 | 249.10 | 259.15 |
| 250.31 | 261.97 | 248.18 | 259.01 |
| 250.25 | 261.66 | 248.93 | 259.03 |
| 250.21 | 261.54 | 249.07 | 259.46 |
| 249.98 | 261.58 | 248.73 | 259.00 |
| 250.21 | 261.30 | 249.52 | 260.16 |
| 250.17 | 261.96 | 248.74 | 258.58 |
| 250.08 | 261.33 | 249.13 | 259.33 |

**XRD, XPS, EDS and Raman spectra**

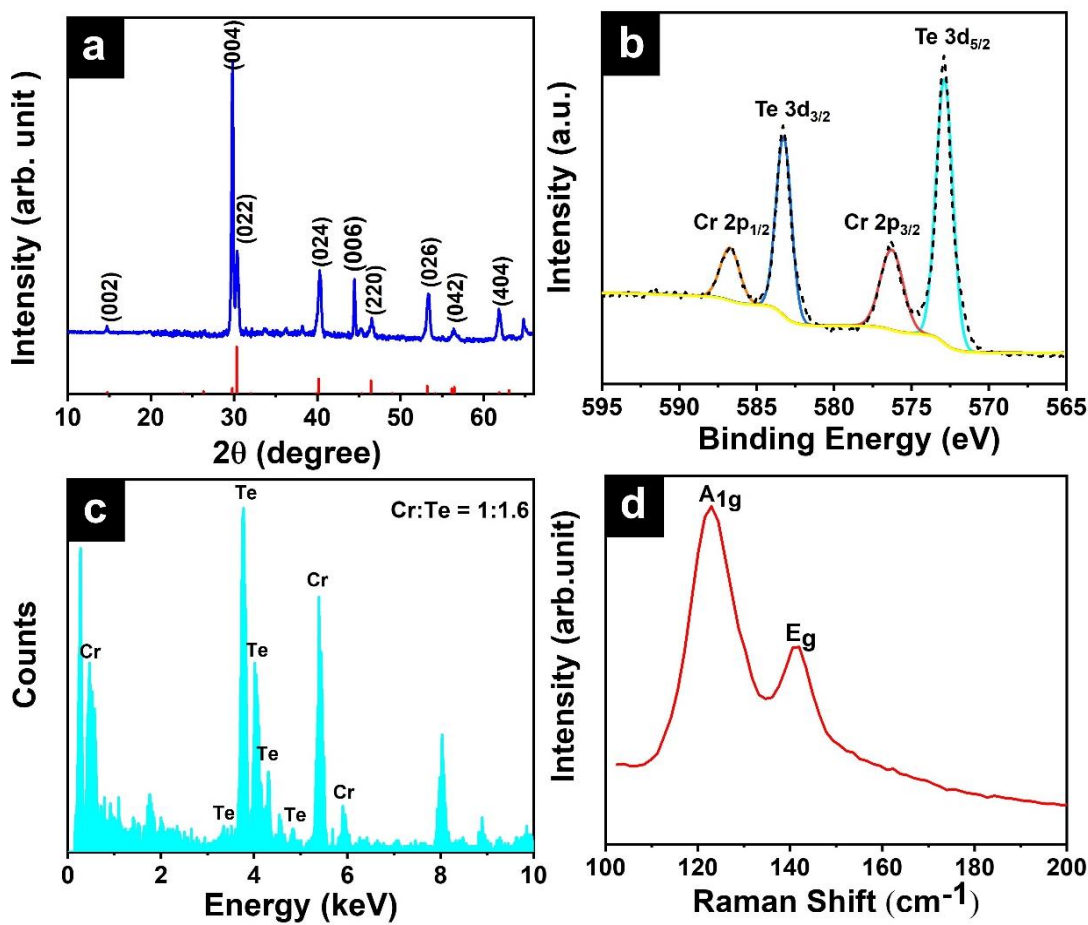

**Figure S8. Compositional and structural characterization of Cr₅Te₈. a,** The XRD spectrum of a Cr₅Te₈ continuous thin film grown at identical conditions to those of 2D crystals; also shown is the standard XRD pattern for hexagonal structured Cr₅Te₈ (PDF#50-1153). **b,** The XPS characterization of as grown Cr₅Te₈ thin film. The peaks located at 586.7 and 576.3 eV are attributed to Cr 2p$_{1/2}$ and Cr 2p$_{3/2}$, while the peaks located at 583.3 and 572.9 eV are attributed to Te 3d$_{3/2}$ and Te 3d$_{5/2}$, respectively. **c,** The EDS spectrum of a single 2D Cr₅Te₈ crystal on WSe₂, which gives a Cr: Te atomic ratio of 1:1.6, consistent with the stoichiomerty of Cr₅Te₈. **d,** Raman spectra of 2D Cr₅Te₈ crystals.

## X-ray absorption spectra (XAS) and XMCD spectra

The XAS and XMCD spectra of a $Cr_5Te_8$ 2D crystal with a thickness of 3 unit cell was measured at the Cr $L_{2,3}$ edges (560.0 – 607.0 eV) at beamline 11.0.1 of the Advanced Light Source at the Lawrence Berkeley National Laboratory. Measurements were conducted at 105 K without an external magnetic field. The x-ray incident angle was 30° with respect to the film plane. Representative XAS with the corresponding XMCD spectra are shown in Figure S9a. To assist the magnetic moment calculations, the integrations of the XMCD and XAS spectra are shown in Figure S9b, c.

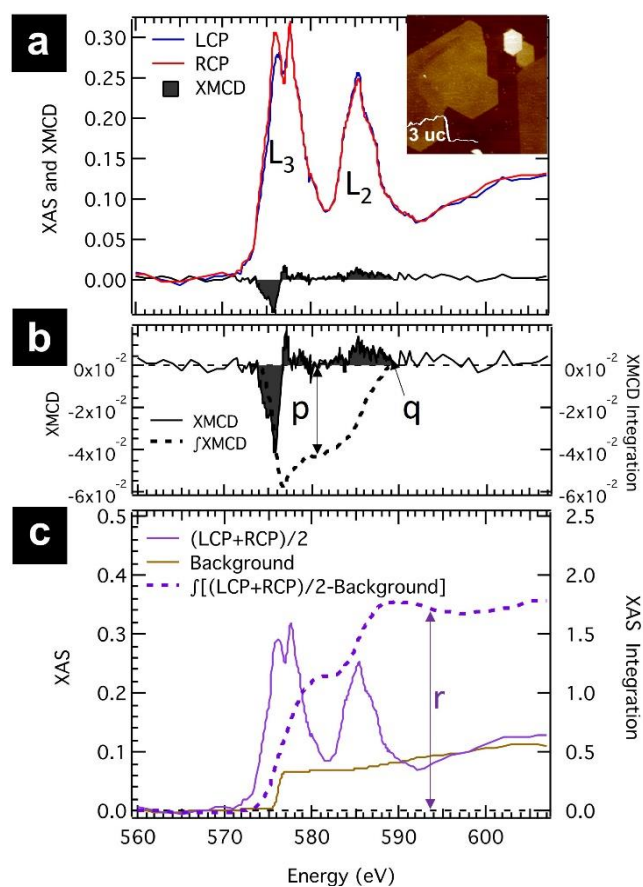

**Figure S9. XAS and XMCD spectra of $Cr^{3+}$ in $Cr_5Te_8$ 2D crystal with 3 unit cell thickness at 105 K. a,** The pair of XAS excited with the left circularly polarized (LCP) and right circularly polarized (RCP) x-rays as well as the resultant XMCD spectrum at the $Cr^{3+}$ $L_{2,3}$. The thickness of the $Cr_5Te_8$ crystal was confirmed by AFM (inset) measurement. **b, c,** The XMCD and the average XAS spectra as well as their integrations. The background line (gold) shown in (**c**) is the two-step-like function that is used for edge-jump removal before the integration.

The spin and orbital magnetic moments per atom, in the units of $\mu_B$/atom, can be determined from the XAS and XMCD spectra in Figure S9 by applying the XMCD sum rules[1], specifically, for 3d transition metal element $Cr^{3+}$ according to the following equations:

$$M_{spin,Cr3+} = -n_{h,Cr3+} \frac{3p-2q}{r} \times SC - <Tz> \qquad (1a)$$

$$M_{orb,Cr3+} = -n_{h,Cr3+} \frac{2q}{3r} \qquad (1b)$$

where $p$, $q$, $r$ are corresponding integrals read from Figure S9b, c for $Cr^{3+}$; the valence hole numbers of $Cr^{3+}$ ($n_{h,\,Cr3+}$) is 7; $SC$ is the spin correction factor estimated for Cr in $Cr_5Te_8$ [2]; $<T_z>$ is the expectation value of the magnetic dipole operator, which is negligible due to the small orbital moment of the 3d element[2-3]. From Figure S9, the calculated $M_{spin,\,Cr3+}$ and $M_{orb,\,Cr3+}$ are 1.2 ± 0.3 $\mu_B$ and -0.02 ± 0.01 $\mu_B$ per Cr atom. Therefore, the sum gives $M_{Cr3+} = 1.2 ± 0.3$ $\mu_B$/atom at 105 K along the x-ray propagation direction, which is 30° away from the sample plane. The comparably large error bar mainly comes from the relatively large uncertainty in absorption background subtraction. Notice that the $Cr_5Te_8$ 2D crystal possesses a perpendicular anisotropy based on RMCD measurements. Therefore, the estimated $M_{spin}$ and $M_{orb}$ should be multiplied by a factor of 2 to project $M_{spin}$ and $M_{orb}$ back to the out-of-plane direction, resulting $M_{spin} = 2.4$ ± 0.6 $\mu_B$/$Cr^{3+}$ and $M_{orb} = -0.04 ± 0.02$ $\mu_B$/$Cr^{3+}$. The estimated spin moment at 0 K is ~ 3 $\mu_B$, which is close to the average $Cr^{3+}$ spin moment of 3.03 $\mu_B$ obtained from DFT calculations.

**Table S2. Bader charges calculated for $Cr_5Te_8$/$WSe_2$ moiré superlattice compared to those of the individual $Cr_5Te_8$ and $WSe_2$ layer (the unit is *e*).**

| | Cr/Cr(int) | Te | W | Se | $Cr_5Te_8$ layer | $WSe_2$ layer |
|---|---|---|---|---|---|---|
| $Cr_5Te_8$/$WSe_2$ | 5.26375/5.2476 | 6.44836 | 5.1464 | 6.4596/ 6.4143 | 701.0056 | 882.9943 |
| $Cr_5Te_8$ | 5.26375/5.32335 | 6.45271 | 5.1410 | | 702 | |
| $WSe_2$ | | | | 6.429 | | 882 |
| Charge transfer | 0/-0.07574 | -0.0044 | 0.0054 | 0.0306/ -0.0147 | -0.9943 | 0.9943 |

In the fifth column, the first number is for the Se atoms at the interface, and the second number is for the atoms at the free surface. The Bader charge analysis shows that there is a charge transfer from $Cr_5Te_8$ towards $WSe_2$. There is ~ 1e transferred across the interface. The largest charge transfer occurs for the intercalating Cr atom. It donates electrons to form a dative bond with Se in $WSe_2$. Some charge is also transferred to W, although significantly smaller. Due to the asymmetry of the charge between Se sites at the interface and free surface, there is a polarity in charge distribution across the $WSe_2$ monolayer.

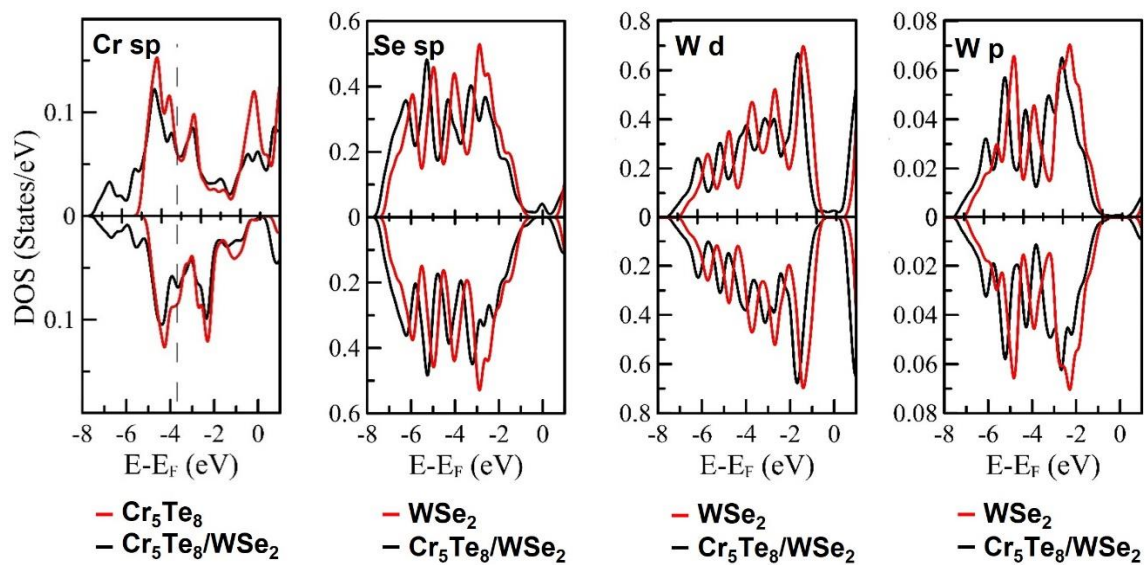

**Figure S10. Site-projected densities of states (DOS) of Cr₅Te₈/WSe₂ moiré superlattice compared to those of individual Cr₅Te₈ and WSe₂ layers.** Left to right: interfacial Cr sp-states, interfacial Se sp-states, W d-states, W p-states. The red curves show DOS of individual layers, and black curves are DOS of Cr₅Te₈/WSe₂ moiré superlattice. DOSs of the representative sites of Cr₅Te₈/WSe₂ moiré superlattice show that there is a redistribution of electron states, most strongly noticeable for the intercalated Cr site. DOSs of W in Cr₅Te₈/WSe₂ superlattice are very similar to those of WSe₂ monolayer, exhibiting a rigid shift of ~ 0.5 eV due to the band alignment across the interface. Fermi energy in the superstructure is close to the conduction band of WSe₂. There is a small charge transfer towards W.